\documentclass[12pt,a4paper]{article}
\usepackage{amsfonts,latexsym}
\usepackage{amsmath,amssymb}
\usepackage{graphicx,color}
\usepackage{multirow}
\usepackage{slashbox}
\numberwithin{equation}{section} \oddsidemargin 0 mm \evensidemargin
0 mm \topmargin -10 mm \textheight 215 mm \textwidth 163 mm

\renewcommand{\thefootnote}{\fnsymbol{footnote}}
\newcommand{\nn}{\nonumber}

\begin{document}
\vspace{12mm}

\begin{center}
{{{\Large {\bf Stability analysis of f(R)-AdS black holes}}}}\\[10mm]

{Taeyoon Moon$^{a}$\footnote{e-mail address: tymoon@sogang.ac.kr},
Yun Soo Myung$^{b}$\footnote{e-mail address: ysmyung@inje.ac.kr},
and
Edwin J. Son$^{a}$\footnote{e-mail address: eddy@sogang.ac.kr}}\\[8mm]

{{${}^{a}$ Center for Quantum Space-time, Sogang University, Seoul, 121-742, Korea\\[0pt]
${}^{b}$ Institute of Basic Sciences and School of Computer Aided Science, Inje University Gimhae 621-749, Korea}\\[0pt]
}
\end{center}
\vspace{2mm}

\begin{abstract}
We study the stability of $f(R)$-AdS (Schwarzschild-AdS) black hole
obtained from  $f(R)$ gravity.  In order to resolve the difficulty
of solving fourth order linearized equations, we transform $f(R)$
gravity into the scalar-tensor theory by introducing two auxiliary
scalars. In this case, the linearized curvature scalar becomes a
dynamical scalaron, showing that all linearized equations are second
order. Using the positivity of gravitational potentials and
$S$-deformed technique allows us to guarantee the stability of
$f(R)$-AdS black hole if the scalaron mass squared satisfies the
Breitenlohner-Freedman bound. This is confirmed by computing
quasinormal frequencies of the scalaron for  $f(R)$-AdS black hole.

\end{abstract}
\vspace{5mm}

{\footnotesize ~~~~PACS numbers: }

\vspace{1.5cm}

\hspace{11.5cm}{Typeset Using \LaTeX}
\newpage
\renewcommand{\thefootnote}{\arabic{footnote}}
\setcounter{footnote}{0}


\section{Introduction}
$f(R)$ gravities~\cite{NO,sf,NOuh} have much attention as one of
strong candidates for explaining the current accelerating
universe~\cite{SN}.  $f(R)$ gravities can be considered as Einstein
gravity  with an additional scalar.   For example, it was shown that
the metric-$f(R)$ gravity is equivalent to the $\omega_{\rm BD}=0$
Brans-Dicke (BD)  theory with a certain potential~\cite{FT}.
However, in order that $f(R)$ gravities are  acceptable, they obey
minimal requirements for theoretical viability~\cite{sf,FT}. Three
important requirements are included: (i) they  possess the correct
cosmological dynamics, (ii) they are free from instabilities of
tachyon and ghost~\cite{PS,CZ,CLF}, (iii) they attain  correct
Newtonian and post-Newtonian limits.

The Schwarzschild-de Sitter black hole was first obtained for a
positively constant curvature scalar in $f(R)$
gravities~\cite{CENOZ} and other black hole solution  was recently
found for a nonconstant curvature scalar in $f(R)$
gravities~\cite{SZ}. A black hole solution was also obtained from
$f(R)$ gravities by requiring the negative constant curvature scalar
$R=\bar{R}<0$~\cite{CDM}. For $1+f'(\bar{R})>0$, this is similar to
the Schwarzschild-AdS (SAdS) black hole obtained from Einstein
gravity. In order to obtain the constant curvature black hole  from
``$f(R)$ gravity coupled to the matter",  the trace of its
stress-energy tensor $T_{\mu\nu}$ should be zero. Two  solutions
were found when the Maxwell~\cite{CDM} and Yang-Mills
fields~\cite{MMS} are coupled to $f(R)$ gravities.

Gravitational stability of a  black hole is a main issue of testing
the adequateness of the solution~\cite{KZ,Vish}. However, the
stability analysis seems not directly applicable  to $f(R)$ black
holes including the Kerr black hole because $f(R)$ gravity contains
fourth-order derivative terms in the linearized
equations~\cite{kerr,BS}. In this case, the requirement (ii)[they
are free from instabilities of tachyon and ghost] would come into
play in testing the stability of $f(R)$ black holes~\cite{FST}. One
may transform $f(R)$ gravity into the scalar-tensor theory to
eliminate fourth-order derivative terms by introducing two scalar
fields~\cite{olmo}. Then, the linearized curvature scalar  became a
scalaron, guaranteeing  that all linearized equations are second
order.  For $f(R)$ black hole, its linearized equations have led to
the same equations for the BD theory~\cite{KKMCP1,KKMCP2}. Using the
stability analysis of black hole in the massive BD
theory~\cite{KKMCP2}, it was  shown that the $f(R)$ black hole is
stable against the external perturbations if the scalaron does not
have a tachyonic mass in asymptotically flat spacetimes~\cite{MMSs}.

 In this work, we investigate the stability of $f(R)$-AdS
(SAdS) black hole.   We  transform $f(R)$ gravity into the
scalar-tensor theory to avoid fourth-order derivative terms by
introducing two auxiliary scalars.  Then, the linearized curvature
scalar  becomes a scalaron, indicating that all linearized equations
are second order. Using the positivity of gravitational potentials
and $S$-deformed technique~\cite{IK} allows us to prove  the
stability of $f(R)$-AdS black hole if the scalaron mass squared
($m_A^2$) satisfies the Breitenlohner-Freedman (BF) bound ($m_A^2\ge
m^2_{\rm BF}=-\frac{9}{4\ell^2}$)~\cite{BF}. In order to confirm the
stability of $f(R)$-AdS black hole, we have to realize that a
practical tool for testing stability in all those cases is a
numerical investigation of quasinormal spectra
$k=k_R-ik_I$~\cite{KZ,BCS,kmsz}. We have checked that there is no
exponentially growing mode (tachyon instability) in anti de Sitter
space if the scalaron mass squared satisfies the
Breitenlohner-Freedman bound ($m^2_A\ge m_{\rm BF}^2$). For
$m^2_A<m^2_{\rm BF}$, however, we have found a growing mode, showing
the tachyonic instability of $f(R)$-AdS black hole.

\section{Perturbation of $f(R)$-AdS black holes}

Let us first consider $f(R)$ gravity without any matter fields
\begin{eqnarray}
S_{f}=\frac{1}{2\kappa^2}\int d^4 x\sqrt{-g} f(R)\label{Actionf}
\end{eqnarray}
with $\kappa^2=8\pi G$. Since all relevant equations derived from
(\ref{Actionf}) are given by Ref.~\cite{MMSs}, we do not reproduce
them here for saving the space. The point to remember is that there
exist fourth-order linearized equations around the $f(R)$-AdS black
holes. In order to avoid it, we introduce  two auxiliary fields
$\phi$ and $A$ which may allow to rewrite the action (\ref{Actionf})
as~\cite{olmo}
\begin{eqnarray}
S_{A}=\frac{1}{2\kappa^2}\int d^4
x\sqrt{-g}\Big[\phi\left(R-A\right)+f(A)\Big]. \label{ActionfA}
\end{eqnarray}
Varying for the fields $\phi$ and $A$ lead to the following
equations:
\begin{eqnarray}
R=A,~~\phi=f'(A),\label{eomA}
\end{eqnarray}
where ${}^{\prime}$ denotes differentiation with respect to its
argument.
 Note that imposing (\ref{eomA}) on the  action (\ref{ActionfA})
recovers  the original action (\ref{Actionf}). On the other hand,
the equation of motion for the metric tensor  can be obtained by
\begin{eqnarray} \label{equa}
f'(A) R_{\mu\nu}-\frac{f(A)}{2}g_{\mu\nu}
+\Big(g_{\mu\nu}\nabla^2-\nabla_{\mu}\nabla_{\nu}\Big)f'(A)=0.\label{eomg}
\end{eqnarray}
In deriving the above equation, we used two relations in
(\ref{eomA}). Considering a constant curvature scalar
$R=\bar{R}=\bar{A}$ together with $\bar{\phi}=f'(\bar{A})={\rm
const}$, Eq.(\ref{eomg}) becomes
\begin{equation}
f'(\bar{A})\bar{R}_{\mu\nu}-\frac{1}{2}\bar{g}_{\mu\nu}f(\bar{A})=0.\label{eomg1}
\end{equation}
Taking the trace of (\ref{eomg1}) leads to
\begin{eqnarray}
\bar{R}=\frac{2f(\bar{A})}{f'(\bar{A})}\equiv 4\Lambda_A.
\label{eqCR}
\end{eqnarray}
which determines the positive, negative and zero curvature scalar by
choosing a specific form of $f(A)$.

 At this stage, we wish to point out the
difference between $R=A$ in (\ref{eomA}) and (\ref{eqCR}). The
former is the strong constraint, while the latter is satisfied only
in the background of constant curvature scalar. In general, $R$ is
determined  by the trace equation
\begin{equation}
R=\frac{-3\nabla^2f'(A)+2f(A)}{f'(A)}.
\end{equation}
Substituting (\ref{eqCR}) into (\ref{eomg1}), one finds the Ricci
tensor which determines the maximally symmetric Einstein spaces
including Minkowski space
\begin{equation}
\bar{R}_{\mu\nu}=\frac{1}{2}\frac{f(\bar{A})}{f'(\bar{A})}\bar{g}_{\mu\nu}=\Lambda_A
\bar{g}_{\mu\nu}.
\end{equation}
A $f(R)$-AdS black hole solution  is given by a spherically
symmetric form
\begin{eqnarray}
 \label{schads}
ds^2=\bar{g}_{\mu\nu}dx^\mu
dx^\nu=-e^{\nu(r)}dt^2+e^{-\nu(r)}dr^2+r^2(d\theta^2+\sin^2\theta
d\varphi^2)
\end{eqnarray}
with
\begin{equation} \label{schnu}
e^{\nu(r)}=1-\frac{2m}{r}-\frac{\Lambda_A}{3}r^2.
\end{equation}
We focus on asymptotically AdS$_4$ spacetime with
$\Lambda_A=-\frac{3}{\ell^2}<0$ and consider the background of
\begin{equation}
\bar{R}=\bar{A}=2f(\bar{A})/f'(\bar{A})<0,
\end{equation}
with $f(\bar{A})<0$.

 Now we introduce the perturbation around the
constant curvature black hole to study stability of the $f(R)$-AdS
black hole
\begin{eqnarray} \label{m-p}
g_{\mu\nu}=\bar{g}_{\mu\nu}+h_{\mu\nu}.
\end{eqnarray}
Hereafter the background quantities are denoted by  the ``overbar''
notation. In this background, we define the perturbations as
\begin{equation}
\bar{R}+\delta R(h)=\bar{A}+\delta A,~~
f(A)=f(\bar{A})+f'(\bar{A})\delta
A,~~f'(A)=f'(\bar{A})+f''(\bar{A})\delta A.
\end{equation}
The first relation implies  the replacement of linearized curvature
scalar by the linearized scalaron as
\begin{equation} \label{rtoa}
\delta R(h)=R-\bar{R}\to \delta A=A-\bar{A}.
\end{equation}
Also the linearized equation to (\ref{equa}) takes the form
\begin{eqnarray}
\delta R_{\mu\nu}(h)&-&\Lambda_A
h_{\mu\nu}+\bar{g}_{\mu\nu}\Bigg[\frac{f''(\bar{A})f(\bar{A})-f'^{2}(\bar{A})}{2f'^{2}(\bar{A})}\bigg]\delta
A \nonumber \\
\label{pertg} &+&\Big[\frac{f''(\bar{A})}{f'(\bar{A})}\Big]
\Big(\bar{g}_{\mu\nu}\bar{\nabla}^2-\bar{\nabla}_{\mu}\bar{\nabla}_{\nu}\Big)\delta
A=0,
\end{eqnarray}
where the linearized Ricci tensor  $\delta R_{\mu\nu}(h)$ is given
by
\begin{eqnarray}
\delta
R_{\mu\nu}(h)&=&\frac{1}{2}\Big(\bar{\nabla}^{\rho}\bar{\nabla}_{\mu}h_{\nu\rho}+
\bar{\nabla}^{\rho}\bar{\nabla}_{\nu}h_{\mu\rho}-\bar{\nabla}^2h_{\mu\nu}-\bar{\nabla}_{\mu}
\bar{\nabla}_{\nu}h\Big).\label{lRmunu}
\end{eqnarray}
It is important to note that taking the trace of $(\ref{pertg})$
with $\bar{g}^{\mu\nu}$ together with (\ref{rtoa}) leads to  the
linearized   ``scalaron" equation as
\begin{eqnarray}
\Big(\bar{\nabla}^2-m_A^2\Big)\delta A=0,\label{gtr}
\end{eqnarray}
where the scalaron mass squared $m_A^2$ is given by
\begin{eqnarray}
m_A^2=\frac{f'^{2}(\bar{A})-2f(\bar{A})f''(\bar{A})}{3f'(\bar{A})f''(\bar{A})}
=\frac{f'(\bar{A})}{3f''(\bar{A})}-\frac{4}{3}\Lambda_A,\label{mass}
\end{eqnarray}
which was already known as (97) of Ref.~\cite{sf} in  de Sitter
spacetimes. Plugging  (\ref{gtr}) into (\ref{pertg}) and rearranging
the terms,  we obtain the linearized  equation
\begin{eqnarray}
\delta R_{\mu\nu}(h)-\Lambda_A
h_{\mu\nu}=\Big[\frac{f''(\bar{A})}{f'(\bar{A})}\Big]\bar{\nabla}_{\mu}\bar{\nabla}_{\nu}\delta
A+\bar{g}_{\mu\nu}\Bigg[\frac{f'^{2}(\bar{A})+f(\bar{A})f''(\bar{A})}{6f'^{2}(\bar{A})}\Bigg]
\delta A. \label{eqRmunu}
\end{eqnarray}
Since the mass dimension of  the  scalaron is two ($[\delta A]=2$),
it would be better to write the canonically linearized equations by
introducing a dimensionless scalaron $\delta \tilde{A}$ defined by
\begin{equation}
\delta \tilde{A}=\frac{f''(\bar{A})}{f'(\bar{A})}\delta A.
\end{equation}
Here $[f'(\bar{A})]=0$ and $[f''(\bar{A})]=-2$.

Finally, we arrive at two linearized equations
\begin{eqnarray}
&&\Big(\bar{\nabla}^2-m_A^2\Big)\delta \tilde{A}=0,\label{gtrf} \\
&& \delta R_{\mu\nu}(h)-\Lambda_A
h_{\mu\nu}-\bar{\nabla}_{\mu}\bar{\nabla}_{\nu}\delta
\tilde{A}-\Big[\frac{m^2_A}{2}+\Lambda_A\Big]\bar{g}_{\mu\nu}\delta
\tilde{A}=0, \label{eqRmunu1f} \end{eqnarray} which should be solved
to carry out the stability analysis of $f(R)$-AdS black hole.

\section{Stability analysis of $f(R)$-AdS black hole}

The metric perturbations $h_{\mu\nu}$ are classified depending on
the transformation properties under parity, namely odd (axial) and
even (polar). Using the Regge-Wheeler~\cite{regge}, and Zerilli
gauge~\cite{Zeri} , one obtains two distinct perturbations : odd and
even perturbations. For odd parity, one has with two off-diagonal
components $h_0$ and $h_1$
\begin{eqnarray}
h^o_{\mu\nu}=\left(
\begin{array}{cccc}
0 & 0 & 0 & h_0(r) \cr 0 & 0 & 0 & h_1(r) \cr 0 & 0 & 0 & 0 \cr
h_0(r) & h_1(r) & 0 & 0
\end{array}
\right) e^{-ikt}\sin\theta\frac{dp_{L}}{d\theta} \,, \label{oddp}
\end{eqnarray}
while for even parity, the metric tensor takes the form with four components $H_0,~H_1,~H_2,$ and $K$ as
\begin{eqnarray}
h^e_{\mu\nu}=\left(
\begin{array}{cccc}
H_0(r) e^{\nu(r)} & H_1(r) & 0 & 0 \cr H_1(r) & H_2(r) e^{-\nu(r)} &
0 & 0 \cr 0 & 0 & r^2 K(r) & 0 \cr 0 & 0 & 0 & r^2\sin^2\theta K(r)
\end{array}
\right) e^{-ikt}p_{L} \,, \label{evenp}
\end{eqnarray}
where $p_L$ is Legendre polynomial with angular momentum $L$ and
$e^{\nu(r)}$ was given by (\ref{schnu}).  In order to achieve  the
stability  of $f(R)$ black hole in asymptotically flat spacetimes,
we have used the result for the stability analysis for the massive
BD theory~\cite{KKMCP1}. However, the present situation is quite
different from $f(R)$ black hole, because we are going to perform
the stability analysis of $f(R)$-AdS black hole.

\subsection{Odd-perturbation}
For odd-parity perturbation, its linearized equation takes a simple
form as
\begin{equation}
\delta R_{\mu\nu}(h)-\Lambda_A h_{\mu\nu}=0.\label{oddeq}
\end{equation}
Using three equations of $\delta R_{t\varphi},~\delta R_{r\varphi},$
and $\delta R_{\theta\varphi}$ in (\ref{oddeq}) together with
(\ref{lRmunu}) and (\ref{oddp}), one finds the following equation
for $h_1$:
\begin{eqnarray} \label{h1eq}
\frac{d^2
h_1}{dr^2}+\left(3\nu^{\prime}-\frac{2}{r}\right)\frac{dh_1}{dr}+
\Bigg[\nu''+2\nu'^2+k^2e^{-2\nu}-e^{-\nu}\frac{(L-1)(L+2)}{r^2}\Bigg]h_1=0.
\end{eqnarray}
Introducing the tortoise coordinate $r^{*}=\int e^{-\nu(r)}dr$ which
maps $r\in[r_+,\infty]$ into  $r^*\in[-\infty,0]$  and a new field
$Q$  defined by
\begin{eqnarray}
Q=e^{\nu}\frac{h_1}{r},
\end{eqnarray}
(\ref{h1eq}) leads to the Regge-Wheeler equation as
\begin{equation}
\frac{d^2Q}{dr^{*2}}+\Big[k^2-V_{RW}\Big]Q=0.\label{oddrw}
\end{equation}
Here the Regge-Wheeler potential is given by~\cite{regge,CL}
\begin{equation}
V_{RW}(r)=\Big(1-\frac{2m}{r}-\frac{\Lambda_A}{3}r^2\Big)\Big[\frac{L(L+1)}{r^2}-\frac{6m}{r^3}\Big].
\end{equation}
It is well known from~\cite{CL} that the effective potential can be
positive definite outside the SAdS black hole when using
$S$-deformed technique~\cite{IK}. This implies that the $f(R)$-AdS
black holes are stable against odd-perturbation because
(\ref{oddrw}) is equivalent to that of the SAdS black hole in
Einstein gravity.

\subsection{Even-perturbation}

We consider the linearized equation (\ref{eqRmunu1f}) fully  to
perform even-parity perturbation. Considering  the metric
perturbation (\ref{evenp}) and the scalar perturbation given by
\begin{equation}
\delta \tilde{A} \propto \Sigma \frac{\psi(r)}{r}
Y^M_{L}(\theta,\varphi) e^{-ikt}\label{scalaro}
\end{equation}
with $Y^M_{L}(\theta,\varphi)$  spherical harmonics, the equation of
either ($\theta\theta$) or ($\varphi\varphi$) component yields
\begin{equation}
H_2=H_0-2\frac{\psi}{r}.
\end{equation}
From five remaining equations of $(tr),~(rr),~(tt),~(t\theta)$, and
$(r\theta)$ components,  one obtains the constraint equation
\begin{eqnarray}
&&\hspace*{-2em}\left\{\tilde{L}^2-2e^{\nu}+re^{\nu}\nu'\right\}H_0+
\left\{2k^2r^2e^{-\nu}+2e^{\nu}+re^{\nu}\nu'+\frac{1}{2}r^2e^{\nu}\nu'^2-\tilde{L}^2+2\Lambda_A
r^2\right\}K\nn\\
&&\hspace*{-2em}-\left\{2ikr+\frac{\tilde{L}^2}{2ik}\right\}H_1+\left\{2\tilde{L}^2-4e^{\nu}-2k^2r^2e^{-\nu}
+\frac{1}{2}r^2e^{\nu}\nu'^2-2\Lambda_A r^2\right\}\frac{\psi}{r}=0
\end{eqnarray}
with $\tilde{L}^2=L(L+1)$.  In deriving this expression, we have
used $\nu''+(\nu')^2+2\nu'/r=-2\Lambda_A e^{-\nu}$. At this stage,
we introduce the tortoise coordinate ($r^{*}$) and a new field
defined by
\begin{equation}
\hat{M}=\frac{1}{pq-h}\left\{p\left(K+\frac{\psi}{r}\right)-\frac{H_1}{k}\right\},
\end{equation}
where
\begin{eqnarray}
q(r)&=&\frac{\lambda(\lambda+1)r^2+3\lambda mr+6m^2}{r^2(\lambda
r+3m)},~~ h(r)=\frac{i(-\lambda r^2+3\lambda
mr+3m^2)}{(r-2m-\frac{\Lambda_A}{3}r^3)(\lambda r+3m)},\nn\\
~~~p(r)&=&-\frac{ir^2}{r-2m-\frac{\Lambda_A}{3}r^3},~~
\lambda=\frac{1}{2}(L-1)(L+2).
\end{eqnarray}
Manipulating  ($tr$) and ($t\theta$)  component equations in
(\ref{eqRmunu1f}), we arrive at the Zerilli's equation\footnote{At
this stage, one may ask the question of ``can one find the single
metric components (e.g. $H_1$ or $\psi$) from the master variable
$\hat{M}$?''.  The answer is obviously ``No''. Actually, one has
$h_{\mu\nu}$ with 10 components in Einstein gravity. A massless
graviton has  2 degrees of freedom (DOF). Upon the Regge-Wheeler
gauge-fixing, one has 6. One DOF  is given by the odd-perturbation
combined by 2($h_1$ and $h_2$) and the other is the
even-perturbation combined by 4 ($H_0,H_1,H_2,K$).  In $f(R)$
gravities, one starts with 11 (10 from $h_{\mu\nu}$ and 1 from
$\psi$), indicating 3 DOF (massless graviton and scalaron) totally.
Hence, the scalaron should be coupled to giving the master variable
$\hat{M}$ in the even-perturbation. }
\begin{equation} \label{evenz}
\frac{d^2\hat{M}}{dr^{*2}}+\Big[k^2-V_{Z}\Big]\hat{M}=0,
\end{equation}
where the Zerilli potential is given by~\cite{Zeri,CL}
\begin{equation}
V_{Z}(r)=\Big(1-\frac{2m}{r}-\frac{\Lambda_A}{3}r^2\Big)
\Bigg[\frac{2\lambda^2(\lambda+1)r^3+6\lambda^2mr^2+18\lambda
m^2 r+18m^3-6\Lambda_Am^2r^3} {r^3(\lambda r+3m)^2}\Bigg].
\end{equation}
The Zerilli potential $V_{Z}$ is always positive for whole range of
$-\infty \le r^* \le 0$ (see Fig. 9 of~\cite{IK}), which implies
that the even-perturbation is stable, even though the scalaron
$\psi(\delta \tilde{A})$ is coupled to making the even-perturbation.
\subsection{Scalar-perturbation}
The main difference between Einstein and $f(R)$ gravities is the
appearance of scalaron. Hence, the stability analysis of scalaron
equation will be considered  to be  the main part of the present
work. In order to obtain the second-order differential equation
for the scalaron, we first recall that the scalaron perturbation
is given by (\ref{scalaro}). In this case, Eq.(\ref{gtrf}) becomes
\begin{equation}
\frac{e^{\nu}}{r}\psi''+\frac{e^{\nu}\nu'}{r}\psi'
-\Bigg[\frac{m_A^2-e^{-\nu}k^2}{r}+\frac{e^{\nu}\nu'}{r^2}+\frac{L(L+1)}{r^3}\Bigg]\psi=0
\end{equation}
Using the tortoise coordinate ($r^{*}$), the above equation reduces
to the Schr\"odinger-type equation
 \begin{equation} \label{scalaron}
\frac{d^2\psi}{dr^{*2}}+\Big[k^2-V_{A}\Big]\psi=0,
\end{equation}
where the scalaron  potential $V_A$ is given by
\begin{equation} \label{sca-pot}
V_{A}(r)=\Big(1-\frac{2m}{r}-\frac{\Lambda_A}{3}r^2\Big)
\Big[\frac{L(L+1)}{r^2}+\frac{2m}{r^3}-\frac{2}{3}\Lambda_A+m^2_A\Big].
\end{equation}
Here, the second term in square bracket is the usual scalar term
with spin zero [in general, $-2m(s^2-1)$ for $s$ spin-weight of a
perturbing field], and  the third  arises  from  AdS$_4$ asymptote.
Importantly, the last term in bracket shows  the feature of a
massive scalaron arisen from $f(R)$ gravity. This term could not be
eliminated using the $S$-deformed technique because it is not an
element of the metric function $e^{\nu(r)}$. Furthermore, it is well
known that there exists the BF bound~\cite{BF} for the Klein-Gordon
type equation (\ref{gtr})
\begin{equation}
m^2_A \ge m^2_{\rm BF}=-\frac{9}{4\ell^2} \end{equation} which
implies an inequality \begin{equation} \label{finequality}
\frac{f'(\bar{A})}{3f''(\bar{A})} \ge -\frac{25}{4\ell^2}.
\end{equation}  On the other hand, we rewrite the
scalaron potential (\ref{sca-pot})
\begin{equation}
V_{A}(r)=\Big(1-\frac{2m}{r}-\frac{\Lambda_A}{3}r^2\Big)
\Big[\frac{L(L+1)}{r^2}+\frac{2m}{r^3}+\frac{6}{\ell^2}+\frac{f'(\bar{A})}{3f''(\bar{A})}\Big],
\end{equation}
which shows  the asymptotic  positivity ($V_A(r\to\infty)\to 0$)
when the following condition is achieved
\begin{equation}
\frac{f'(\bar{A})}{3f''(\bar{A})}\ge - \frac{6}{\ell^2}.
\end{equation}
Hence, we allow  the negative mass squared  in  AdS$_4$ spacetimes
which belongs to the region
\begin{equation}
-\frac{25}{4\ell^2} \le \frac{f'(\bar{A})}{3f''(\bar{A})} \le -
\frac{6}{\ell^2} \to -\frac{9}{4\ell^2} \le m^2_A \le
-\frac{2}{\ell^2}. \label{mass-b}\end{equation} We note that the
left bound comes from the BF bound, while the right bound arises
from the  asymptotic positivity condition of $V_A(r\to\infty)\to0$.
Accordingly, we divide $m^2_A$ into four regions\footnote{The fourth
region $m^2_A\ell^2 \ge 0$ corresponds to ``tachyon-free condition"
required  by the stability of de Sitter space of
$(f'^2-2ff'')/3f'f'' \ge 0$~\cite{CZ,CLF,fn,olmod} and Minkowski
space of $f'/3f''\ge0$~\cite{olmo}. This implies that $f'>0$ and
$f''>0$ for stability. However,  the stable region (tachyon-free
case) could be extended to negative mass squared  in the AdS space:
$m^2_A \ell^2 \ge -9/4[(\ref{finequality})]$, which implies that
$f''(\bar{A})<0$, for $f'(\bar{A})>0$ in AdS space. This is in
contrast with other perturbation analysis: the Dolgov-Kawasaki
instability with $f''<0$ in cosmological perturbations~\cite{DK},
graviton and scalar propagations in the
Minkowski~\cite{olmo,olmof}.}: $m^2_A\ell^2<m^2_{\rm
BF}\ell^2=-9/4,~-9/4\le m^2_A\ell^2\le-2,~-2< m^2_A\ell^2<0,$ and
$m^2_A\ell \ge 0$. The boundaries are $m^2_A\ell^2=0,-2,-9/4$. As is
shown in Fig. 1, there are four types of graphs including three
boundaries, depending on the mass squared $m^2_A$.  This may imply
that there is no ghost instability even for the negative mass
squared is allowed, if $m^2_A$ satisfies the bound (\ref{mass-b}).
This is surely a new feature of stability condition for $f(R)$-AdS
black holes.  On the other hand, we expect that for $m^2_A<m^2_{\rm
BF}$, there will be the tachyonic instability.
\\
\begin{figure*}[t!]
   \centering
   \includegraphics{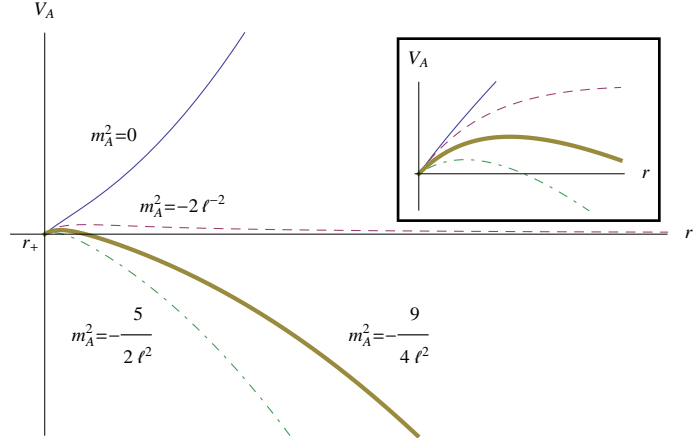}
\caption{Scalaron potential  $V_A$ graphs as function of $r$ with
fixed $\ell^2=1$: for $m^2_A\ell^2=0$ case, it shows the massless
scalaron in AdS spacetimes, while for $m^2_A\ell^2=-2$ case, it
approaches zero as $r\to\infty$ (asymptotic positivity). For two
cases of $m^2_A\ell^2=-9/4$ and $-5/2$, their potentials approach
negative infinity as $r\to \infty$.} \label{VA}
\end{figure*}

\section{Quasinormal modes}

It is well  known that a practical tool for testing stability of
all kinds of black holes is a numerical investigation of the
quasinormal spectra~\cite{KZ}. Hence, we wish to confirm the
stability of $f(R)$-AdS black holes by computing quasinormal
modes.
Considering  $\Psi(t,r_*)=e^{-ikt}\Psi(r_*)$, the boundary
conditions for quasinormal modes propagating in the $f(R)$-AdS
black hole background are given by
\begin{equation}
\Psi(r_*) \sim \left\{ \begin{aligned} & e^{-ikr_*}, & & \text{for }
r_* \to -\infty, \\ & 0, & & \text{for } r_* \to 0,
\end{aligned} \right. \label{qnmbc}
\end{equation}
where $\Psi(r_*)$ represents $Q$ (odd), $\hat{M}$ (even), and $\psi$
(scalaron) in \eqref{oddrw}, \eqref{evenz}, and \eqref{scalaron},
respectively.  The frequencies of the quasinormal modes are given by
complex numbers, $k=k_R-ik_I$, and the modes with $k_I>0$ ($k_I<0$)
corresponding to decaying (growing) ones, since the time dependence
 is given by $e^{-ikt}=e^{-k_It}e^{-ik_Rt}$.  We note that
if there is at least one growing mode, the black hole  is
unstable.

Now,~(\ref{oddrw}), (\ref{evenz}), and (\ref{scalaron}) could  be
written as
\begin{equation}
\frac{d^2\Psi(r_*)}{dr_*^2} + (k^2-V) \Psi(r_*) = 0 \label{qnmeq}
\end{equation}
with $V$ denoting $V_{RW}$, $V_{Z}$, and $V_{A}$.  Multiplying this
equation by a complex conjugated function $\bar\Psi(r_*)$ and
integrating it, we have
\begin{align}
I =& \int^{0}_{-\infty} dr_* \left[ \bar\Psi(r_*) \frac{d^2\Psi(r_*)}{dr_*^2} + (k^2-V) \left| \Psi(r_*) \right|^2 \right] \notag \\
=& ik \left| \Psi(-\infty) \right|^2 
+ \int^{0}_{-\infty} dr_* \left[ - \left| \frac{d\Psi(r_*)}{dr_*}
\right|^2 + (k^2-V) \left| \Psi(r_*) \right|^2 \right]= 0,
\label{inteq}
\end{align}
where we used the boundary condition~\eqref{qnmbc}. The imaginary
part of \eqref{inteq} takes the form
\begin{equation}
\text{Im}(I) = k_R \left[ \left| \Psi(-\infty) \right|^2 - 2 k_I
\int^{0}_{-\infty} dr_* \left| \Psi(r_*) \right|^2 \right] = 0,
\end{equation}
which implies that either $k_R=0$ or $k_I>0$, so that the growing
modes cannot oscillate  as was  pointed out in~\cite{kmsz,KZ}.
That is, the unstable mode is defined by the condition of
\begin{equation}
{\rm unstable~mode} \to k_R=0,~~k_I<0 \label{unstablec}.
\end{equation}
in the quasinormal-mode approach.
 Next, the real part of \eqref{inteq} is given by
\begin{align}
\text{Re}(I) =& k_I \left| \Psi(-\infty) \right|^2 + \left( k_R^2 - k_I^2 \right) \int^{0}_{-\infty} dr_* \left| \Psi(r_*) \right|^2 \notag \\
  & - \int^{0}_{-\infty} dr_* \left[ \left| \frac{d\Psi(r_*)}{dr_*} \right|^2 + V \left| \Psi(r_*)\right|^2 \right] =
  0,
\end{align}
which means  that the unstable mode  does not exist if the potential
$V$ is positive definite. We confirm that  there is no unstable
modes for the even-perturbation since $V_{Z}$ is always positive.
For the odd-perturbation, $V_{RW}$ becomes negative near the
horizon, but the stability may be  achieved by the $S$-deformation
technique. On the other hand, it turned out that there is no growing
modes in odd- and even-quasinormal mode computations~\cite{CL,CKL}.
Hence, we do not need to compute quasinormal frequencies for the
metric perturbations again.

 However, we remind the reader  the instability of the scalar
perturbation depends on  the scalaron mass squared, $m_A^2$.
Requiring $V_A$ to be positive corresponds to  the bound of
$m_A^2\ge-2\ell^{-2}$.  Thus, one expects that there is no
exponentially growing mode for this case. Also, if $m^2_A$ satisfies
the BF bound ($m^2_{\rm BF} \le m^2_A \le -2\ell^{-2}$), there is no
tachyon instability. Finally, if $m^2_A<m^2_{\rm BF}$, there will be
tachyonic instability. In order to show it explicitly, we calculate
quasinormal frequencies of the scalaron numerically.

To this end, Eq.~\eqref{qnmeq} should be  rewritten in the ingoing
Eddington coordinates with $v=t+r_*$, following the original work
of Ref.\cite{HH} for large Schwarzschild-AdS black holes:
\begin{equation} \label{lqnm}
e^\nu \frac{d^2\tilde\Psi}{dr^2} + \left( \nu' e^\nu - 2 i k \right)
\frac{d\tilde\Psi}{dr} - \tilde{V} \tilde\Psi = 0,
\end{equation}
where $\Psi(t,r_*) = e^{-ikv} \tilde\Psi$ and $V = e^\nu
\tilde{V}$. Then, changing coordinate variable $r$ into
$x=\ell/r$, the equation (\ref{lqnm}) becomes
\begin{equation}
s(x) \frac{d^2\tilde\Psi}{dx^2} + \frac{t(x)}{x-x_+}
\frac{d\tilde\Psi}{dx} + \frac{u(x)}{(x-x_+)^2} \tilde\Psi = 0,
\end{equation}
where
\begin{align}
s(x) =& \frac{x^4e^\nu}{x_+-x} = \frac{x^2}{x_+^3} \left[ x_+^2 + x_+x + (1+x_+^2)x^2 \right] = \sum_{n=0}^4 s_n (x-x_+)^n, \\
t(x) =& -2i\tilde{k}x^2 - 2x^3 + \frac{3}{x_+^3}(1+x_+^2)x^4 = \sum_{n=0}^4 t_n (x-x_+)^n, \\
u(x) =& (x-x_+)\tilde{V} = \sum_{n=0}^4 u_n (x-x_+)^n.
\end{align}
Here $x_+=\ell/r_+$ and $\tilde{k} = k\ell$, and the boundary
condition~\eqref{qnmbc} takes the form of  $\tilde\Psi(x\to0)\to0$.
Considering a series solution
\begin{equation}
\tilde\Psi = \sum_{n=0}^\infty a_n (x-x_+)^n,
\end{equation}
the boundary condition at infinity yields an algebraic equation
\begin{equation}
\sum_{n=0}^\infty a_n(-x_+)^n = 0.
\end{equation}
Solving this, one could  find (discrete) quasinormal modes.
\begin{table}[tdp]
\caption{\label{tab:qnm}Frequencies of quasinormal modes of large
$f(R)$-AdS black holes for the scalaron}
\begin{center}
\begin{tabular}{|c|c|c|c|c|c|c|}
  \hline
  \multirow{2}*{\backslashbox{$m_A^2\ell^2$}{$r_+/\ell$}} 
  & \multicolumn{2}{c|}{$10$} & \multicolumn{2}{c|}{$50$} & \multicolumn{2}{c|}{$100$} \\
  \cline{2-7}
  & $\tilde{k}_R$ & $\tilde{k}_I$ & $\tilde{k}_R$ & $\tilde{k}_I$ & $\tilde{k}_R$ & $\tilde{k}_I$ \\
  \hline
  0 
  & $18.6070$ & $26.6418$ & $92.4937$ & $133.193$ & $184.953$ & $266.386$ \\
  $-1/4$ 
  & $18.0708$ & $25.6542$ & $89.8263$ & $128.258$ & $179.619$ & $256.515$ \\
  $-1/2$ 
  & $17.500$ & $24.6056$ & $86.988$ & $123.018$ & $173.945$ & $246.036$ \\
  $-3/4$ 
  & $16.888$ & $23.4827$ & $83.941$ & $117.407$ & $167.85$ & $234.813$ \\
  $-1$ 
  & $16.222$ & $22.2661$ & $80.629$ & $111.327$ & $161.23$ & $222.653$ \\
  $-5/4$ 
  & $15.486$ & $20.9261$ & $76.968$ & $104.630$ & $153.91$ & $209.260$ \\
  $-3/2$ 
  & $14.64$ & $19.42$ & $72.81$ & $97.067$ & $145.6$ & $194.13$ \\
  $-7/4$ 
  & $13.66$ & $17.63$ & $67.88$ & $88.15$ & $135.7$ & $176.3$ \\
  $-2$ 
  & $12.3592$ & $15.3307$ & $61.4106$ & $76.6646$ & $122.797$ & $153.330$ \\
  $-9/4$ 
  & $9.9$ & $12$ & $49$ & $58$ & $98$ & $116$ \\
  $-5/2$ 
  & $0$ & $-8.64855$ & $0$ & $-32.5878$ & $0$ & $-59.5519$ \\
  $-11/4$ 
  & $0$ & $-52.5789$ & $0$ & $-225.272$ & $0$ & $-467.935$ \\
  $-3$ 
  & $0$ & $-126.996$ & $0$ & $-536.309$ & $0$ & $-1035.61$ \\
  \hline
\end{tabular}
\end{center}
\end{table}
In Table~\ref{tab:qnm}, we list  quasinormal modes for the scalaron
($L=0$) obtained by truncating the series after 140 terms.
The first row of $m^2_A\ell^2=0$ indicates the quasinormal mode for
the massless scalar, which is consistent with Ref.~\cite{HH}. The
last three rows in Table 1 shows clearly that there exist unstable
modes of $k_I<0$ with $k_R=0$ for the scalaron mass
($m_A^2\ell^2=-5/2<-9/4$) below the BF bound. This corresponds to
the tachyonic instability. For $-9/4\le m^2_A \ell^2 \le -2$, there
is no unstable mode, showing that there is no tachyon instability.
This confirms the presence of  the BF bound even for $f(R)$-AdS
black holes.

\section{Discussions}

We have investigated the stability of $f(R)$-AdS (Schwarzschild-AdS)
black hole obtained from the $f(R)$ gravity. Even though the
Schwarzschild-AdS black hole is known to be stable in Einstein
gravity, checking the stability of $f(R)$-AdS black hole is a
nontrivial task because the linearized Einstein equation is fourth
order in $f(R)$ gravity. In order to resolve this difficulty, we
have translated the fourth-order equation into the second-order
equation by introducing auxiliary scalar fields $A$ and $\phi$.
 In this case, the linearized curvature scalar  $\delta R(h)$ becomes a massive scalaron $\delta A$, showing that all linearized equations are second
 order.

 The stability on the metric perturbations remains
 unchanged,  confirming that the odd (even) perturbations lead to
 the Schr\"odinger-type equation with the Regge-Wheeler (Zerilli)
 potentials in asymptotically AdS$_4$ spacetimes. Actually, this corresponds to the case in Einstein gravity, even though
 the even-perturbation contains the scalaron in addition to $H_0,~H_1,~H_2,$ and
 $K$. For the odd-perturbation, we have needed the $S$-deformation
 method to prove the stability of $f(R)$-AdS black hole because the corresponding potential becomes  negative near the event horizon.

The main difference comes from the linearized scalaron equation
 because the scalaron plays the role of  a massive scalar, which is physically  propagating on the  $f(R)$-AdS black
 hole  background. We have shown that $f(R)$-AdS black hole is
 stable against the scalaron-perturbation   if the scalaron
mass squared $m^2_A$ satisfies the BF bound. Especially for $-9/4\le
m^2_A \le -2$, there is no unstable mode (exponentially growing
mode), showing that there is no tachyon instability.   This confirms
the presence of the BF bound even for $f(R)$-AdS black holes.  This
is consistent with graviton and scalar propagations
AdS~\cite{Myungf} spacetimes, but contrasts to other perturbation
analysis: the Dolgov-Kawasaki instability with $f''(R)<0$ in
cosmological perturbations~\cite{DK}, graviton and scalar
propagations in the Minkowski~\cite{olmo,olmof}, and de Sitter
spacetimes~\cite{fn,olmod}.

Finally, we have found the tachyonic instability for $m^2_A<m^2_{\rm
BF}$.  This is confirmed by computing quasinormal frequencies of the
scalaron for large $f(R)$-AdS black hole. The last three rows in
Table 1 shows clearly that there exist unstable modes of $k_I<0$
with $k_R=0$ for the scalaron mass ($m_A^2\ell^2=-5/2<-9/4$) below
the BF bound. This indicates the tachyonic instability.

 \vspace{1cm}

{\bf Acknowledgments}

 T. Moon and E. Son were supported by the
National Research Foundation of Korea (NRF) grant funded by the
Korea government (MEST) through the Center for Quantum Spacetime
(CQUeST) of Sogang University with grant number 2005-0049409. Y.
Myung  was supported by the National Research Foundation of Korea
(NRF) grant funded by the Korea government (MEST) (No.2011-0027293).

\end{document}